# Discrepant hardening observed in cosmic-ray elemental spectra

H. S. Ahn[1], P. Allison[2], M. G. Bagliesi[3], J. J. Beatty[2], G. Bigongiari[3], J. T. Childers[4], N. B. Conklin[5], S. Coutu[5], M. A. DuVernois[4], O. Ganel[1], J. H. Han[1], J. A. Jeon[6], K. C. Kim[1], M. H. Lee[1], L. Lutz[1], P. Maestro[3], A. Malinin[1], P. S. Marrocchesi[3], S. Minnick[7], S. I. Mognet[5], J. Nam[6], S. Nam[6], S. L. Nutter[8], I. H. Park[6], N. H. Park[6], E. S. Seo[1,9*], R. Sina[1], J. Wu[1], J. Yang[6], Y. S. Yoon[1,9], R. Zei[3], and S. Y. Zinn[1]

[1]*Institute for Physical Science and Technology University of Maryland, College Park, MD 20742, USA*

[2]*Department of Physics, Ohio State University, Columbus, OH 43210, USA*

[3]*Department of Physics, University of Siena and INFN, Siena 53100, Italy*

[4]*School of Physics and Astronomy, University of Minnesota, MN 55414, USA*

[5]*Department of Physics, Penn State University, University Park, PA 16802, USA*

[6]*Department of Physics, Ewha Womans University, Seoul 120-750, S. Korea*

[7]*Department of Physics, Kent State University Tuscarawas, New Philadelphia, OH 44663, USA*

[8]*Department of Physics and Geology, Northern Kentucky University, Highland Heights, KY 41099, USA*

[9]*Department of Physics, University of Maryland, College Park, MD 20742, USA*

---

[*] Corresponding author, seo@umd.edu

**ABSTRACT**

The balloon-borne Cosmic Ray Energetics And Mass (CREAM) experiment launched five times from Antarctica has achieved a cumulative flight duration of about 156 days above 99.5% of the atmosphere. The instrument is configured with complementary and redundant particle detectors designed to extend direct measurements of cosmic-ray composition to the highest energies practical with balloon flights. All elements from protons to iron nuclei are separated with excellent charge resolution. Here we report results from the first two flights of ~70 days, which indicate hardening of the elemental spectra above ~200 GeV/nucleon and a spectral difference between the two most abundant species, protons and helium nuclei. These results challenge the view that cosmic-ray spectra are simple power laws below the so-called "knee" at $\sim 10^{15}$ eV. This discrepant hardening may result from a relatively nearby source, or it could represent spectral concavity caused by interactions of cosmic rays with the accelerating shock. Other possible explanations should also be investigated.

# 1. INTRODUCTION

The transport of cosmic rays through Galactic magnetic fields randomizes their arrival directions and obscures their sources. Supernova shock waves can provide the power required to sustain the intensity of these energetic particles, but there are many open questions about the details of the acceleration mechanism. The current paradigm for their origin in supernovae is based on a steady state/continuous source distribution that results in a simple power law for all elements. The true source distribution is more likely discrete in time and space, and structures ("bumps") in energy spectra could reflect a non-uniform distribution of cosmic-ray sources. More recent sources would dominate the high-energy spectra, and this is where the effect of discreteness in time is greatest (Taillet et al. 2004). Electrons lose their energy rapidly via synchrotron radiation and inverse Compton scattering, which leads to a spectral cutoff above $\sim 10^{12}$ eV. Observation of higher energy cosmic-ray electrons at Earth means that their sources must be near us in space and time (Kobayashi et al. 2004). Local sources might also be detected as spectral changes at lower energies, as has been proposed to explain recent electron and positron observations (Chang et al. 2008; Adriani et al. 2009; Abdo et al. 2009). Likewise, nearby sources might also be reflected in the spectra of nuclei. The magnitude of a bump from a discrete source with respect to nuclei background should be less prominent than an electron bump, since nuclei spectra steepen mainly through the diffusive propagation mechanism rather than radiative energy loss.

Cosmic rays entering the atmosphere interact with atmospheric nuclei to produce secondary particles that can reach the ground if the incident energy is high enough. Cosmic rays in the energy range $10^{14} - 10^{20}$ eV have been detected via ground-based observations of the particle showers they initiate in the atmosphere. These measurements have shown that the all-particle spectrum has features known as the "knee" and "ankle" corresponding, respectively, to regions of spectral steepening at $\sim 10^{15}$ eV and flattening at $\sim 10^{18}$ eV. The cosmic-ray spectrum has otherwise been believed to follow a smooth, grand power law of $\sim E^{-2.7}$. The substantial contribution of a nearby and recent single source (supernova remnant or pulsar) to the flux of protons and nuclei has been proposed (Erlykin & Wolfendale 1999) to explain the "knee."

Ground-based measurements provide the large collecting power needed to observe the rapidly decreasing cosmic-ray flux with increasing energy, but they cannot unambiguously identify the primary particle that initiated the shower. Direct measurements with satellite or balloon-borne detectors can identify the primary particle and determine its energy, although the energy reach is currently limited to $\sim 10^{15}$ eV by the detector size and exposure time. The latter have provided primary cosmic-ray energy spectra with rather good precision at energies up to $\sim 10^{11}$ eV/nucleon (e.g., Engelmann et al. 1990; Müller et al. 1991; Aguilar et al. 2002; Haino et al. 2004). The pioneering direct measurements above this energy with balloon-borne emulsion chambers show large discrepancies and uncertainties (Asakimori et al. 1998; Derbina et al. 2005). Consequently, the exact shape of the elemental spectra, e.g., whether or not the spectral index is the same for all elements, including protons, has remained a tantalizing question. Precise measurements of the energy dependence of elemental spectra from $\sim 10^{12}$ to $\sim 10^{15}$ eV, where the expected rigidity-dependent supernova acceleration limit could be reflected in a composition change, provide a key to understanding cosmic-ray acceleration and propagation.

The Cosmic Ray Energetics And Mass (CREAM) investigation (Seo et al. 2008) was conceived to measure the detailed energy dependence of elemental spectra to the highest energy possible with a balloon-borne instrument. The goal was to understand the acceleration and Galactic propagation of the bulk of cosmic rays. That included whether and how the "knee" structure in the all-particle spectrum observed by air shower experiments is related to the mechanisms of acceleration, propagation, and confinement.

The CREAM project has had five successful flights at float altitudes between ~38 and ~40 km. The balloons were launched from McMurdo, Antarctica, and each flight subsequently circumnavigated the South Pole two or three times. We report here results from the CREAM-I and –II flights of 42 days and 28 days, respectively. The data from subsequent flights are still being analyzed. These direct measurements bridge the energy gap between lower-energy direct measurements and the abundant indirect measurements at higher energies from the ground (e.g., Antoni et al. 2002).

## 2. EXPERIMENTAL PROCEDURE

The CREAM instrument shown schematically in Fig. 1 is configured with redundant and complementary charge identification and energy measurement systems. Starting at the top, they include a Timing Charge Detector (TCD); a Transition Radiation Detector (TRD) flown on CREAM-I but not CREAM-II; a Cherenkov Detector (CD); and a calorimeter module consisting of a Silicon Charge Detector (SCD), carbon targets, scintillating fiber hodoscopes (S0/S1 and S2), and an ionization calorimeter (W-scn) comprised of a stack of tungsten plates with interleaved scintillating-fiber layers. Details of the detectors and their performance are discussed elsewhere (Ahn et al. 2007a). The TCD defines the 2.2 $m^2$ sr trigger geometry and measures the incident particle charge using fast electronics before backscattered particles hit the detector (Ahn et al.

2009a).  The CD vetoes low-energy background particles due to the low geomagnetic cutoff over Antarctica.  The TRD determines the Lorentz factor of $Z \geq 3$ nuclei and measures the rise of the ionization signal in the proportional tubes for low energies. The SCD is segmented into ~2 cm$^2$ pixels to minimize hits of accompanying backscattered particles in the same segment as the incident particle.  The carbon target induces hadronic interactions in the calorimeter, which consists of stacked layers of tungsten interleaved with scintillating fiber ribbons.  The calorimeter measures the shower energy and provides tracking information to determine which segment(s) of the charge detectors to use for the charge measurement.  The scintillating fiber ribbons sample the energy deposited by the showers, and they provide 3-D imaging of the compressed shower development in the dense tungsten absorber.

The instrument employs about 10,000 electronic channels to readout the highly segmented detectors. It was calibrated pre-flight at the European Organization for Nuclear Research (CERN) using the highest-energy proton and electron test beams available.  As discussed in Yoon et al. (2005), Park et al. (2004), Marrochesi et al. (2005) and Yoon et al. (2007), the particle beam data are in good agreement with detailed Monte Carlo simulations. The instrument was also exposed to A/Z = 2 nuclear fragments of the 158 GeV/nucleon indium beam at CERN. The energy deposit as a function of mass number shows good linearity for 158 GeV to ~9 TeV incident energy (Ahn et al. 2006).  Our simulations show that the calorimeter response is quite linear and that its resolution is nearly energy independent up to $10^{15}$ eV, where the experiment is limited by low particle fluxes (Ahn et al. 2001). Particle energies were determined from ionization energy deposits of cascades initiated in the calorimeter. The calorimeter energy de-convolution included corrections for both the small energy dependence of shower leakage and the energy resolution, as described in Ahn et al. (2009b).  A substantial fraction of the $Z \geq 3$ cosmic rays were measured in both the calorimeter and TRD, thereby providing direct in-flight cross-calibration of their energy measurements (Maestro et al. 2007). The TRD analysis and measurement of the secondary-to-primary (e.g., B/C) ratio have been reported elsewhere (Ahn et al. 2008).

The trajectory of each event was reconstructed from a linear fit to the core of the shower axis through the multiple layers of scintillating fiber strips in the calorimeter (Zei et al. 2007; Ahn et al. 2007b). The extrapolation of this reconstructed trajectory was required to traverse the active areas of both the Silicon Charge Detector and the bottom of the calorimeter. The signal in the silicon pixel selected for the charge measurement was corrected for the angle of incidence before making the charge determination. As shown in Fig. 2, individual elements were identified with excellent, $\sigma \sim 0.2e$ charge resolution.

The measured spectra were corrected for attenuation due to interactions in the air above the balloon altitude (3.9 g/cm$^2$ on average) and background from misidentified charges (e.g., 5% for protons and 7% for helium). The latter takes into account interactions above the charge detector and the effect of particles backscattered from the calorimeter. The absolute flux was obtained by correcting the measured spectra for the trigger, reconstruction and event selection efficiencies (~70 %), the geometry factor (e.g., 0.46 m$^2$-sr for protons and helium) and live time (~56% and ~75 %, respectively, for CREAM-I and CREAM-II). Statistical uncertainties were estimated with 84% Poisson confidence limit for the highest energy bins, where the number of particles is less than 10. Uncertainties in the nucleus-nucleus charge-changing cross sections used to correct for interactions in the instrument and atmosphere contribute ~2% to the flux uncertainty. The systematic uncertainty in the energy scale is estimated to be ~ 5%. Considering all the uncertainties in the instrument efficiencies, the overall systematic uncertainties in the absolute fluxes are estimated to be ~10 %. Overall systematic uncertainties may shift the spectra up or down, but they would not affect the spectral shapes.

## 3. RESULTS

Our spectra at the top of the atmosphere from $2.5 \times 10^3$ GeV to $2.5 \times 10^5$ GeV can be represented by power-law fits (flux $\propto E^\beta$) with indices $\beta$ of -2.66 ± 0.02 for protons and -2.58 ± 0.02 for helium, respectively. These spectra in energy per particle are compared with previous low-energy measurements in Fig. 3, where the fluxes are multiplied by $E^{2.75}$ to facilitate visual comparison with the lower energy measurements. Specifically, extrapolation of the Alpha Magnet Spectrometer (AMS) spectra with indices of -2.78 ± 0.009 for protons and -2.74 ± 0.01 for helium (Alcaraz et al. 2000) would appear as nearly horizontal lines. Our proton and helium spectra are both harder (flatter) than the lower energy measurements. Our helium fluxes are 4 standard deviations higher than would be indicated by extrapolation of a single power-law fit of the AMS helium data to our measurement range. Our significantly lower proton-to-helium ratio of 8.9 ± 0.3 at ~ 9 TeV/nucleon compared to the 18.8 ± 0.5 ratio estimated from the AMS fluxes at 100 GeV/nucleon verifies that the proton spectrum is not parallel to the helium spectrum. The AMS ratio agrees with the CAPRICE ratio (~ 18) and BESS ratio (~16) at similar energies.

Whether or not the proton spectrum index is the same as that of heavier nuclei has long been a tantalizing question. It has been difficult to prove this subtle difference, because spectral indices determined from measurements over the limited energy range of a single experiment could not provide a definitive answer. Our measurements over a wide energy range at high energies, where no solar modulation effect is expected, show this difference clearly.

The CREAM helium and heavier nuclei spectra are shown as functions of energy per nucleon and compared with previous measurements in Fig. 4. Here the observed fluxes are multiplied by $E^{2.5}$, so the high-energy spectra will appear nearly horizontal to facilitate visual comparison among the elements. These compiled data show similar spectral shapes with a bump around 10 – 20 GeV/nucleon where the effect of solar modulation becomes negligible. They also show a harder spectrum for each element above ~200 GeV/nucleon, indicating departure from a single power law. Our helium fluxes are slightly lower than the fluxes reported by the Advanced Thin Ionization Calorimeter (ATIC-2), but both CREAM and ATIC-2 measurements show harder spectra than the lower energy measurements (Alcaraz et al. 2000). Our fluxes are consistent with the pioneering measurements of the Japanese-American Cooperative Emulsion Experiment – JACEE (Asakimori et al. 1998) above ATIC energies, but they are higher than the Russian Nippon Joint Balloon (RUNJOB) data (Derbina et al. 2005).

The CREAM C – Fe data are consistent with the HEAO-3 (Engelmann et al. 1990) and CRN (Müller et al. 1991) data at low energies, and the TRACER (Ave et al. 2008) data where they overlap. We note that there is only one TRACER data point between ~10 GeV/nucleon and ~400 GeV/nucleon, where we observe spectral shape changes. A single-power law fit to our data agrees with the TRACER O – Fe power-law fit (Ahn et al. 2009b), but the data above 200 GeV/nucleon tend to be systematically higher than a single power-law fit indicates. A broken power law gives a better fit to our data. Note that the JACEE and RUNJOB experiments did not report spectra of individual elements heavier than helium.

Considering the limited statistics, we investigated broken power law fits with the spectral indices $\gamma_1$ and $\gamma_2$, respectively, below and above 200 GeV/nucleon. Within the current statistics, the fits and their significance are nearly the same for any breakpoint in the range 200 – 250 GeV/nucleon. The broken power-law fit for elements heavier than Carbon were normalized to the Carbon fit. The resulting fit indices shown in Fig. 5 are $\gamma_1 = -2.77 \pm 0.03$ and $\gamma_2 = -2.56 \pm 0.04$, which differ by 4.2 σ. The spectral index $\gamma_1$ is consistent with the low energy helium measurements, e.g., the AMS index of $-2.74 \pm 0.01$, whereas $\gamma_2$ agrees remarkably well with our CREAM helium index of $-2.58 \pm 0.02$ at higher energies. We note that the experiment-to-experiment index variations for the low energy data are slightly larger than their quoted fit errors, probably due to different energy ranges for their fits and residual effects of solar modulation.

## 4. DISCUSSION

An explanation for the difference between proton and helium spectra could be that they are coming from different types of sources or acceleration sites. For example, protons might come mainly from the supernova explosion of a low mass star directly into the interstellar medium. Helium and heavier nuclei might come mainly from the explosion of a massive star into the atmosphere swept out by the progenitor star rather than directly into the general interstellar medium (Biermann 1993). The strong stellar wind of the massive star would be magnetic and enriched by mass ejections that expose its deeper layers. The acceleration rate could be determined at first by the magnetic field of the progenitor's wind, which might be significantly higher than the magnetic field in the interstellar medium. In this case, the resulting spectra of helium and heavier nuclei from the wind would be harder than the spectrum of protons originating from a low-mass star explosion into the interstellar medium.

The spectral hardening observed above ~200 GeV/nucleon could result from a nearby isolated supernova remnant, or it could be the effect of distributed acceleration by multiple remnants embedded in a turbulent stellar association (Medina-Tanco & Opher 1993). Most massive stars are born in associations, and they evolve quickly enough to explode as supernovae in the vicinity of their parent molecular cloud. The dynamic effect of repeated supernova explosions in a small region of the Galaxy is large bubbles – superbubbles – of hot material surrounded by a shell of compressed interstellar matter. Superbubbles powered by fast stellar winds and clusters containing hundreds of massive stars, called OB associations, have been proposed as the acceleration site for Galactic cosmic rays (Axford 1981; Butt & Bykov 2008). Furthermore, observations of isotopic and elemental abundances of heavy and ultra-heavy nuclei support the concept of cosmic-ray acceleration in OB associations (Binns et al. 2007).

Alternatively, the source spectra could be harder than previously thought based on the low energy data, or the hardening could reflect the predicted concavity in the spectra before the "knee" (Hillas 2005). In the framework of diffusive shock acceleration, cosmic-ray pressure created by particle interactions with the shock could broaden the shock transition region, causing higher energy particles to gain energy faster. This could result in spectral flattening with increasing cosmic-ray energy and deviations from a pure power law (Ellison et al. 2000). The observable effect is expected to be small when summed over multiple sources and propagated over Galactic distances (Allen et al. 2008), but the possible observation of concavity would provide evidence that cosmic rays are dynamically important in the acceleration process.

Our results impact interpretation of various experimental observations, including ground-based air-shower measurements that rely on hadronic interaction models to interpret their results. If the observed spectral hardening is due to concavity, it could be indicating that our energy is approaching the "knee" in the all-particle spectrum. If the hardening is from a local astrophysical source, the acceleration limit of that source could cause the "knee." In addition, changes in the conventional propagation and acceleration model that account for spectral hardening of nuclei would impact the search for dark matter annihilation products by refining the cosmic-ray background level.

## 5. CONCLUSIONS

The confluence of precise cosmic ray measurement capabilities and Antarctic long-duration balloon flights near the top of the atmosphere going multiple times around the South Pole are providing new clues for understanding cosmic rays. The CREAM data presented herein clearly show the subtle difference in protons and helium spectra. The helium spectrum agrees well with the spectra of heavier nuclei from carbon to iron, and discrepant hardening of all the observed spectra is evident above ~200 GeV/nucleon. The coincidence of this observed hardening at a rigidity similar to electron enhancements reported earlier (Chang et al. 2008) indicates that a single mechanism might be responsible for all the elements, as well as electrons. Whatever the explanation, our results contradict the traditional view that a simple power law can represent cosmic rays without deviations below the "knee" at $\sim 10^{15}$ eV. The pervasive discrepant hardening in all of the elemental spectra we have observed provides important constraints on cosmic ray acceleration and propagation models, and it must be accounted for in an explanation of the mysterious cosmic ray "knee."

## ACKNOWLEDGMENTS

The authors thank the NASA Wallops Flight Facility Balloon Program Office, Columbia Scientific Balloon Facility, National Science Foundation Office of Polar Programs, and Raytheon Polar Services Company for the successful balloon launches, flight operations, and payload recoveries. This work was supported in the U.S. by NASA grants NNX08AC11G, NNX08AC15G, NNX08AC16G and their predecessor grants, in Italy by INFN, and in Korea by the Creative Research Initiatives of MEST/NRF.

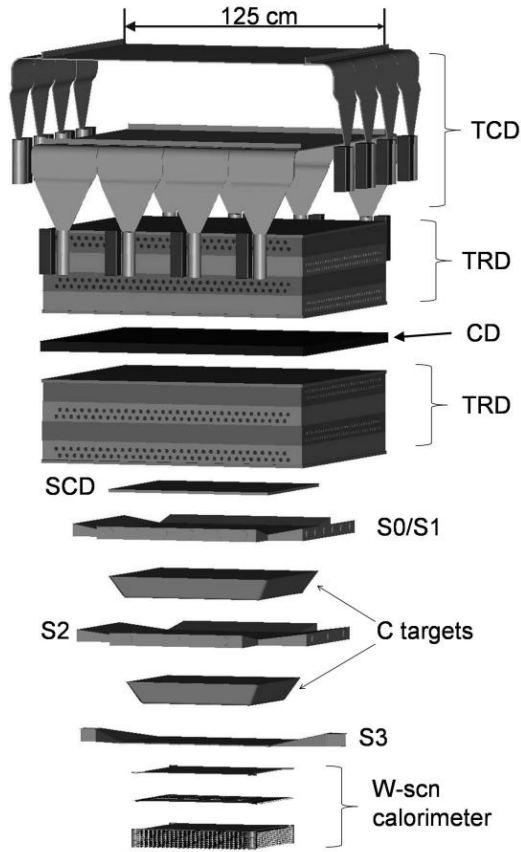

Fig. 1. Schematic of the CREAM-I instrument configuration. The TRD and calorimeter provide complementary energy measurements, as well as in-flight cross-calibration of their energy scales using particles heavier than helium measured in both detectors. Tracking for showers is accomplished by extrapolating each shower axis in the calorimeter back to the charge detectors. Hodoscopes in the carbon target provide additional tracking information above the tungsten stack. The TRD provides tracking for particles that do not interact above the calorimeter.

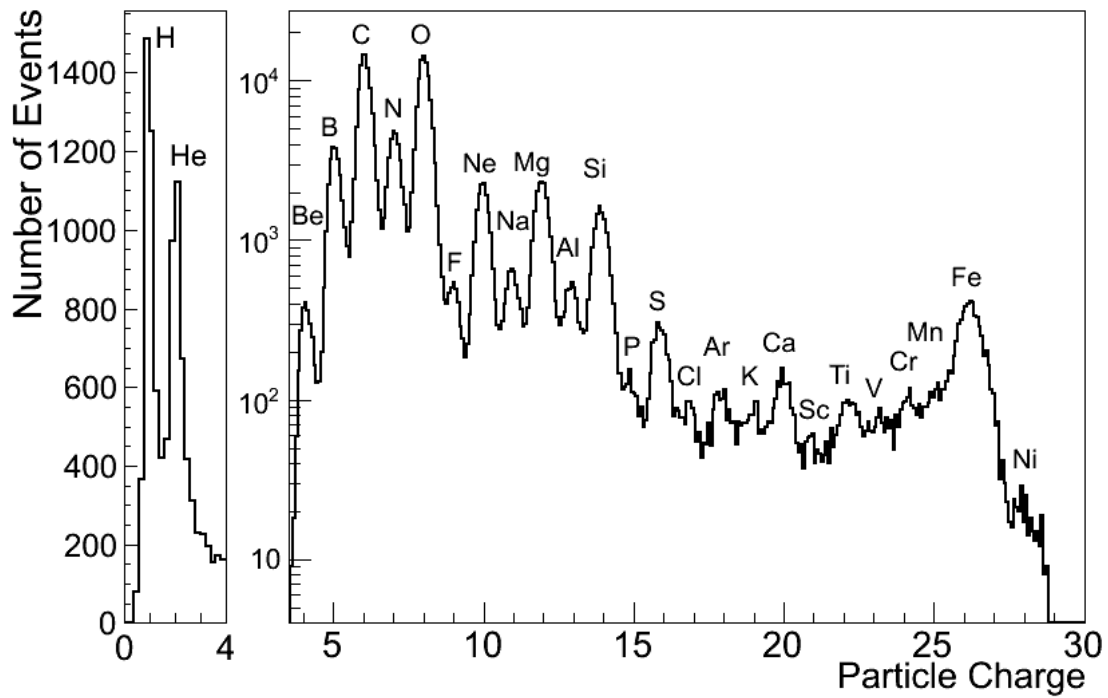

Fig. 2. Distribution of cosmic-ray charge measured with the SCD. The charge reconstructed for a fraction of the flight data is shown in units of the elementary charge e. The individual elements are clearly identified. The charge resolution is better than 0.2e for proton and helium, 0.2e for oxygen, and slightly worse than 0.2e for higher charges. The relative abundance in this plot has no physical significance, because needed corrections for interactions and propagations have not been applied to these data.

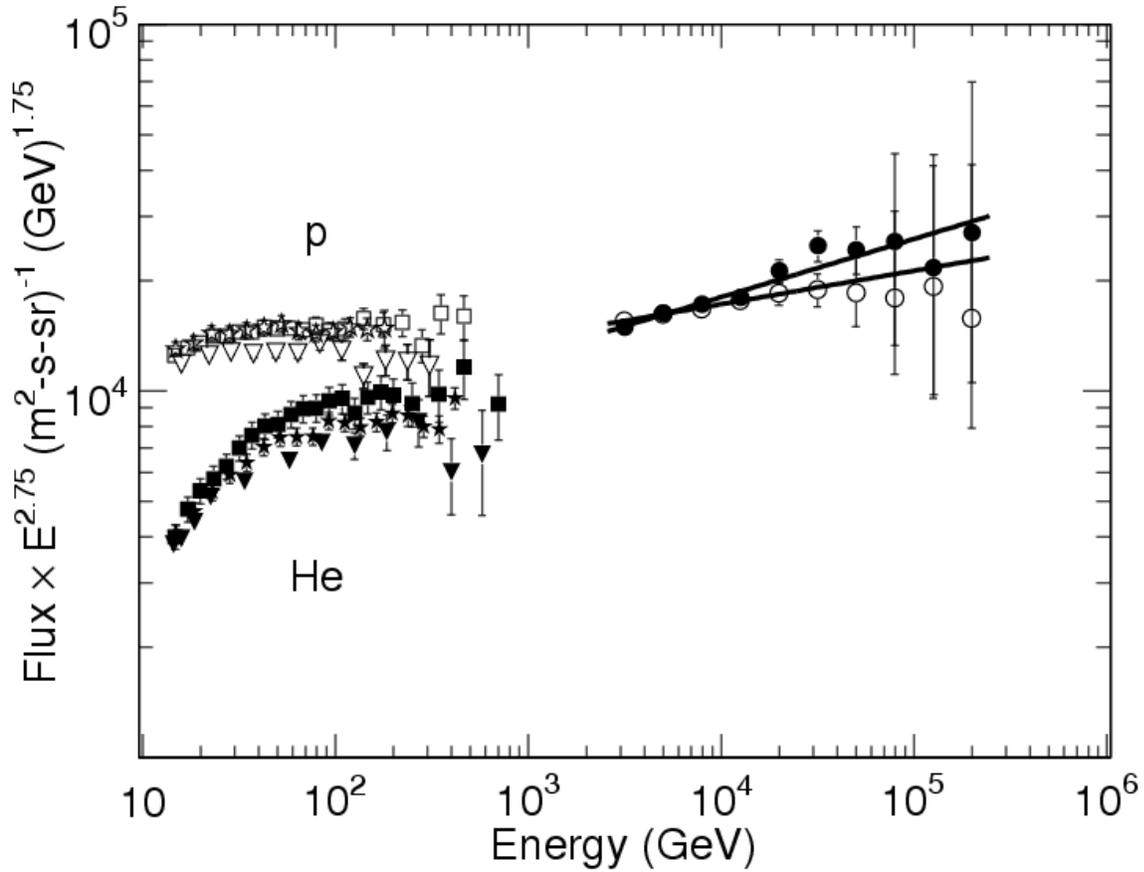

Fig. 3. Measured energy spectra of cosmic-ray protons and helium nuclei. The CREAM-I spectra are compared with selected previous measurements (Alcaraz et al. 2000; Haino et al. 2004; Boezio et al. 2003) using open symbols for protons and filled symbols for helium: CREAM (circles), AMS (stars), BESS (squares), CAPRICE (inverted triangles). The error bars represent one standard deviation, which is not visible when smaller than the symbol size. The lines represent power-law fits to the CREAM data.

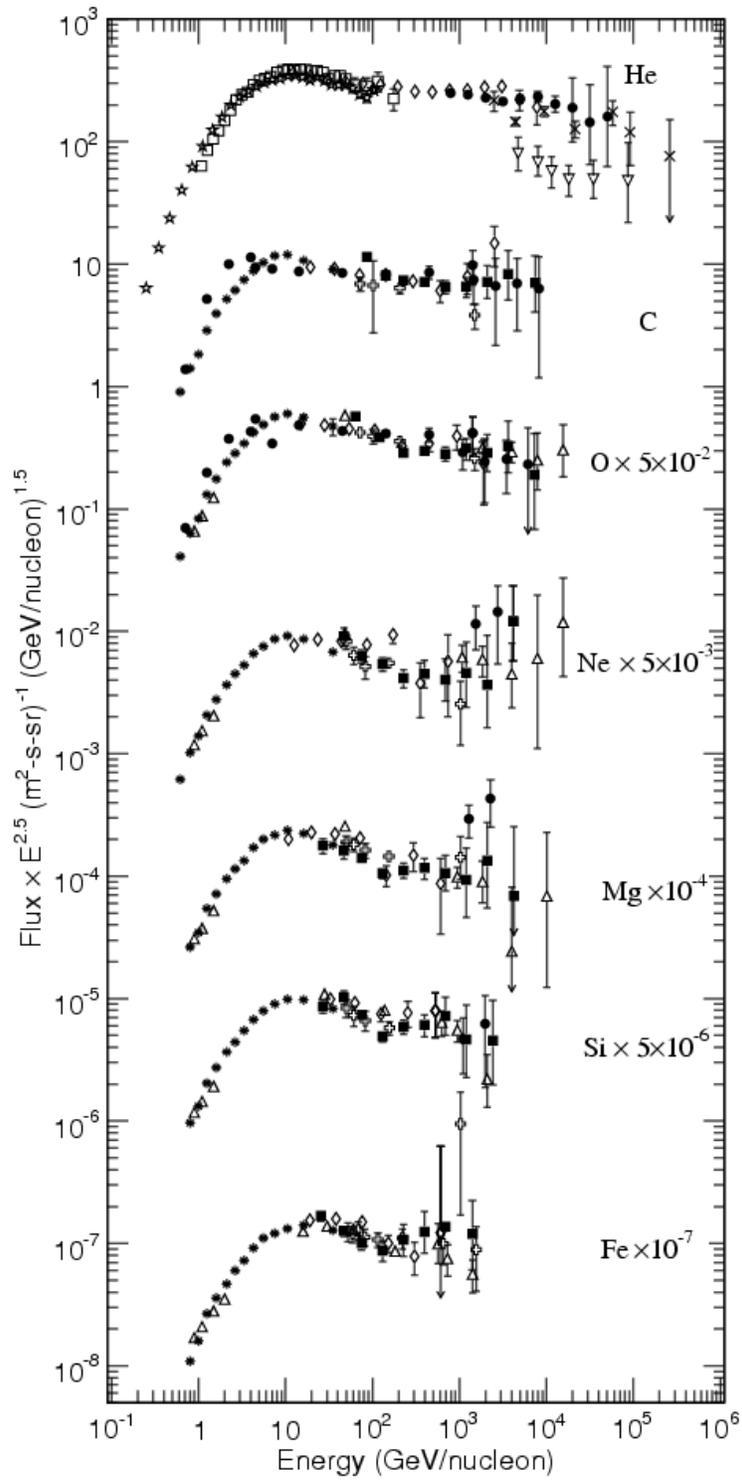

Fig. 4. Compilation of helium and heavier nuclei data. The CREAM elemental fluxes are compared with selected previous data (Asakimori et al. 1998; Derbina et al. 2005; Zei et al. 2007; Ahn et al. 2007b; Alcaraz et al. 2000; Panov et al. 2009): CREAM-1 (filled circles), CREAM-2 (filled squares), AMS (stars), BESS (open squares), JACEE (X), RUNJOB (inverted triangles), HEAO-3 (asterisks), CRN (open crosses), TRACER (triangles), and ATIC-2 (diamonds). The data for elements heavier than C were multiplied by the indicated factors to separate their fluxes in the figure. The error bars represent one standard deviation, which is not visible when smaller than the symbol size.

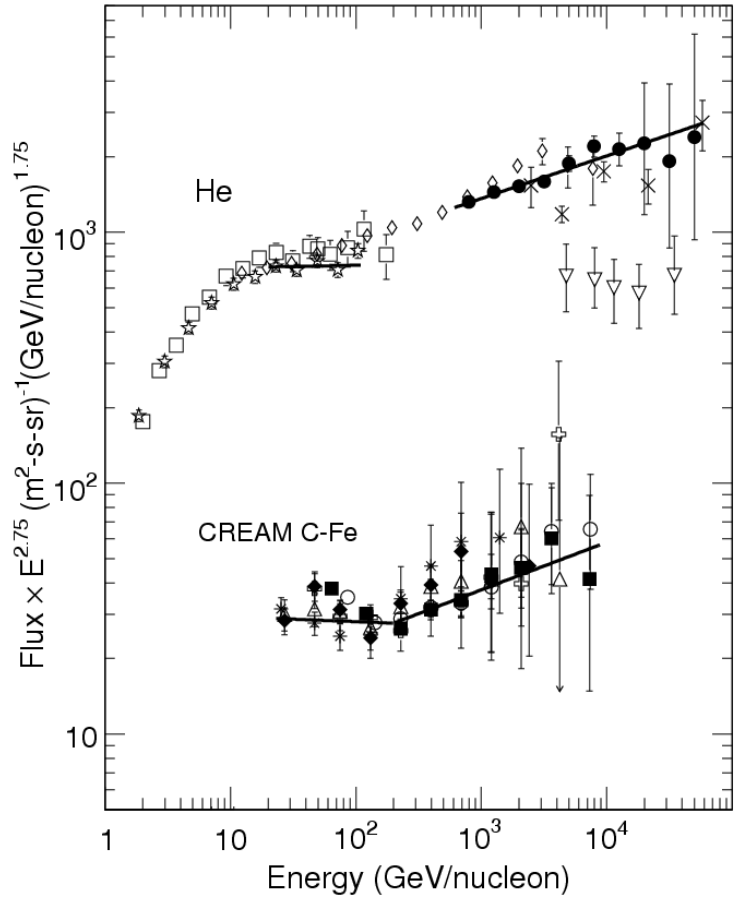

Fig. 5. Broken power-law fit to helium and heavier nuclei data. The lines for helium represent a power-law fit to AMS (open stars) and CREAM (filled circles) data, respectively. Also shown are helium data from other experiments: BESS (open squares), ATIC-2 (open diamonds), JACEE (X), and RUNJOB (open inverted triangles). Some of the overlapping BESS and AMS data points are not shown to achieve better clarity. The lines for C-Fe data represent a broken power-law fit to the CREAM heavy nuclei data: Carbon (open circles), Oxygen (filled squares), Neon (open crosses), Magnesium (open triangles), Silicon (filled diamonds), and Iron (asterisks).